The Rendering from the Periodic System of the Elements on the stability, elastic, and electronic properties of $M_2AC$ phases

Erxiao Wu a,b,c,1, Yiming Zhang a,c,1, Mian Li a,c, Youbing Li a,c, Kan Luo a, Shiyu Du a,c,*, Qing Huang a,c,**

*a*Engineering Laboratory of Advanced Energy Materials, Ningbo Institute of Materials Technology and Engineering, Chinese Academy of Sciences, Ningbo 315201, China.
*b*University of Chinese Academy of Sciences, Beijing 100049, China.
*c*Qianwan Institute of CNiTECH, Ningbo 315460, China.

ABSTRACT

MAX phases are nanolaminated ternary materials that combine metallic and ceramic properties. Currently, the A-site elements replacement in traditional ones by later transition-metals opens a door to explore new types of MAX phases. In this work, we performed a systematic first-principle study to explore trends in stability, electronic structure and mechanical properties of 288 compositions of $M_2AX$ phases (M=Ti, V, Cr, Zr, Nb, Mo, Hf, Ta, W; A=Al, Si, P, S, Ga, Ge, As, Se, In, Sn, Sb, Te, Tl, Pb, Bi, Mn, Fe, Co, Ni, Cu, Zn, Tc, Ru, Rh, Pd, Ag, Cd, Os, Ir, Pt, Au, Hg; X=C). Such a dataset, combined with the rigid-band model been applied to most transition metal carbides, shows us the fundamental trends in bonding mechanisms and mechanical properties of MAX phases endowed with the periodic arrangements of M/A-site elements. It worth noting, in particular, the M-A *d-d* interactions of MAX phases uniquely contribute to the elastic constant $C_{33}$.

1. Introduction

The $M_{n+1}AX_n$ (n = 1~3) phases are nanolaminated ternary carbides and nitrides with hexagonal crystal structure (space group $P6_3/mmc$); where "M" belongs early transition metals, "A" is an element mostly from group 13 or 14, and X is C and/or N [1, 2]. The combination of both metallic and ceramic properties brings about their potential applications as high-temperature electrodes [3], components with resistance to friction and wear [4], structural material in extreme environment [5-8]. The discoveries of in-plane solid-solutions and out-of-plane ordered phases open the doors for their family expansion. More fascinating, MAX phases play indispensable roles in 'top-down' synthesis of their 2D counterparts, MXenes; which have shown their potentials for prospective applications [9, 10]. At present, a large number of MAX phases, with different types of M and A elements, have been experimentally reported or theoretically predicted; continually broadening their family members, as well as opening pathways to explore novel MAX phases [11-17].

Due to their chemical diversity, as well as tunable physical and chemical properties; it deserves to explore new types of MAX phases that beyond the currently known configurations. Owing to time and cost constraints, it is impractical to investigate the synthesizability of all possible MAX phases, as well as their corresponding properties, using experimental techniques. In contrast, theoretically to initially predict the formability of potential MAX phases and their properties, and subsequently, to provide guidance for the experimental synthesis is quite feasible. During the past decade, high-throughput computation strategy [18-21] has been introduced in for the explorations of potential MAX phases. Cover *et al.* [4] carried out computational survey of the elastic properties of 240 elemental combinations using first-principles calculations; and revealed the governing role of the A element and its interaction with the M element on the *c* axis compressibility and shearability of the material, as well as the minor role of the X element. Keast *et al.* [22] utilized density-functional theory (DFT) to study the stability of five distinct systems (Ti-Al-C, Ti-Si-C, Ti-Al-N, Ti-Si-N, and Cr-Al-C, with n = 1~4) by comparing the total energies to that of competing phases. Dahlqvist *et al.* [23] studied the stability trends of MAX phases with M = Sc, Ti, V, Cr, Mn including a set of known most competing phases; and suggest that the maximum stability is reached for V and Ti as a transition metal for C- and N-containing MAX phases, respectively. Aryal *et al.* [24] performed the calculations on 792 different MAX phases and found that 665 are thermodynamically and elastically stable. Ashton *et al.* [25] employed DFT-based approach to calculate the formation energies of 216 pure $M_2AX$ pahses and 10314 solid solutions relative to their competing phases. Ohmer *et al.* [26, 27] investigated stability and physical properties of the MAX phases with the chemical formula $M_2AX$, with the "M" element being one of the 3*d*, 4*d*, or 5*d* transition metal elements, the "A" element being a main group element, and X being either C or N. Recently, Khaledialidusti *et al.* [28] studied the synthesizability of a series of MAX phases (M = Sc, Ti, V, Cr, Y, Zr, Nb, Mo, Hf, Ta, W; A = Al, Si, P, S, Cu, Zn, Ga, Ge, As, Cd, In, Sn, Ir, Au, Tl, Pb, Bi) and to investigate their feasible exfoliation to generate 2D systems from high-throughput DFT calculations.

*Corresponding author.
**Corresponding author.
  E-mail addresses: dushiyu@nimte.ac.cn (S. Du), huangqing@nimte.ac.cn (Q. Huang).

In this work, we aim to fully explore the potential 211-MAX phases built from the periodic table as well as the stability, elastic, and electronic properties of potential M$_2$AC configurations (M = Ti, V, Cr, Zr, Nb, Mo, Hf, Ta, W; A = Al, Si, P, S, Ga, Ge, As, Se, In, Sn, Sb, Te, Tl, Pb, Bi, Mn, Fe, Co, Ni, Cu, Zn, Tc, Ru, Rh, Pd, Ag, Cd, Os, Ir, Pt, Au, Hg), which are investigated via high-throughput DFT calculations (Fig. 1). By employing the rigid-band theory [29], the transformations of bonding characteristics are expounded for the shifting of A-site from main group to transition metal group. The heat of formation (HoF) for the MAX candidates, $E_{form}^{MAX}$, are evaluated by using the total energy of the bulk MAX structure, as well as the total energies (per atom) of the M, A, and X elements in their standard bulk phases. Here, the thermodynamic stabilities of the MAX phases are evaluated *via* Hof, rather than a more rigorous but far more time consuming one based on thermodynamic assessment on all potential competing phases in the M-A-X ternary phase diagrams, treated as a reasonable compromise [30]. It is found that the bonding integrity of MAX phase is modulated by both of *valence electron concentration* (VEC) and *geometrical* factor, where the effect of VEC can be counteracted by the *geometrical geometrical* factor through the filling of M-A *d-p* and *d-d* orbitals. The heatmaps of calculated HoF and mechanical properties for all MAX candidates with respect to M- and A-site elements display apparent periodic trends; revealing the effects of Periodic Law of elements on the formability and their properties of MAX phases. It is worth noting that the aim of this work is to figure out the underlying mechanisms which interpret the trends of the formability, as well as electronic and elastic properties, of MAX phases; thus further to attempt to order the thermodynamic and kinetic stabilities of MAX phase without cover-all calculations carried out.

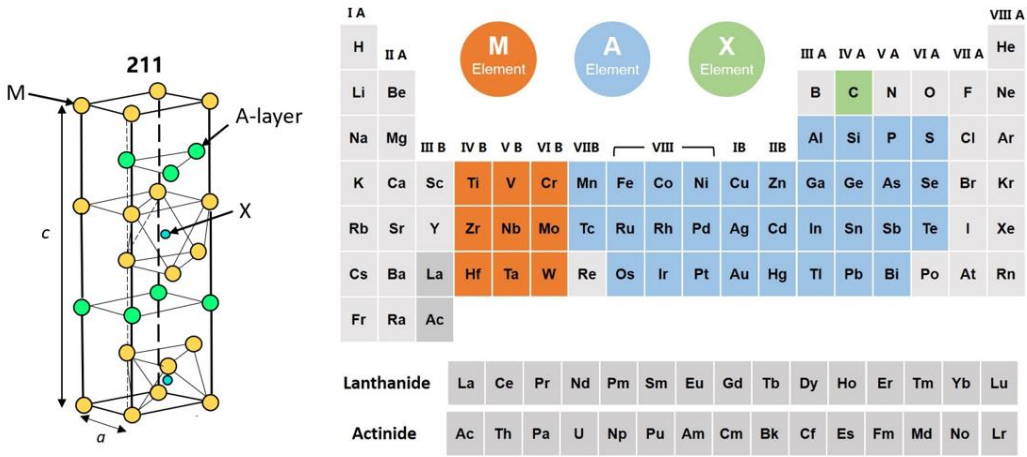

**Fig. 1.** Schematic of the chemical search space within this work.

## 2. Calculation Details

All the first-principles density functional calculations are performed using the Vienna Ab initio Computing Software Package [31, 32], and the projector-augmented wave method [33] with an energy cutoff of 500 eV is adopted. For the exchange and correlation functionals, we use the Perdew–Burke–Ernzerhof (PBE) version of the generalized gradient approximation (GGA) [34].

Based on the conjugate gradient scheme [35], all the structures are relaxed until the forces on each atom are less than $1.0 \times 10^{-4}$ eV/Å, and the total energy is converged at $1.0 \times 10^{-8}$ eV/cell. A Γ-centered *k*-point mesh of $12 \times 12 \times 4$ is adopted for describing the Brillouin zones (BZs) of 3D MAX phases configurations. The Gaussian smearing with a broadening of 0.01 eV is employed to plot the density of states. All the volume, lattice parameters, and atomic positions are allowed to relax for the optimization of bulk states. For crystals with hexagonal symmetry, elastic constants are calculated based on the stress–strain relationship, and then the bulk (K), shear (G) and Young's modulus (E) are obtained from the Voigt–Reuss–Hill approximation.[36, 37] And there are reference formulas for other mechanical performance parameters [38].

All the COHP calculations are done using the Local Orbital Basis Suite Towards Electronic-Structure Reconstruction (LOBSTER) code [39-41] with the pbeVaspFit2015 and koga basis set, and all the charge spilling are less than 5% [41]. All the structures are visualized in the VESTA code [42].

The formation energy and exfoliation energy are used to assess the difficulty of its generation and etching, which are calculated from Eqs. (1) [4, 24, 43].

$$E_{form}^{MAX} = (E_{tot}^{MAX} - n_M E_{tot}^M - n_A E_{tot}^A - n_X E_{tot}^X)/(n_M + n_A + n_X) \qquad (1)$$

where $n_M$, $n_A$, and $n_X$ are the number of M, A, and X atoms, respectively, in the bulk MAX structure. The monatomic energies of standard bulk phases are from the open quantum materials database (The Materials Project) [44].

## 3. Results and Discussions

## 3.1 The investigation of formability and electronic properties

In order to investigate the formability of MAX candidates, the ground-state properties are investigated at first. The optimized lattice constants and their HoF are shown as heatmaps in Fig. 2 a-c. It is obvious that the lattice parameter $a$ of MAX phases with M in the same period increases significantly with the arrangement of groups (the increasing of electron). For lattice parameter $c$, their values increase along the Group of M; when the A-site element changes within the same period, the values increase first and then decrease, where the maximum value of $c$ approach to MAX phases with A elements possessing the highest number of valence shell electrons within the period. For the HoF: 1) The values rise gradually with the arrangement of the M-site element groups (the increase of electrons) in the same period; 2) it augment gradually with the increase of the periods. In previous studies, we have pointed out the important role of *geometrical* factor and *electron concentration factor* of MAX phases for guiding their synthesis [45]. Integrating above views with the perspectives raised from other literature [24, 38, 46-48], the above trends can be interpreted through the rigid-band model.

MAX phases are one type of materials conforming to the rigid-band model, where the increasing of VEC would not change the shape of their density of states; only the Fermi energy level moves up and down. Therefore, this mechanism has been employed to study the bonding laws at the Fermi energy level and to infer the characteristics and trends of MAX phases. When the VEC reaches 8.4, like $V_2AlC$ (Fig. 3a), the hybrid bonding of M-A $p\&d_{t_{2g}}-p_z$ reaches saturation [45]. When the VEC exceeds 8.4, such as $Cr_2AlC$ (Fig. 3b), the anti-bond of $d$-$d$ metal fills below the Fermi level.

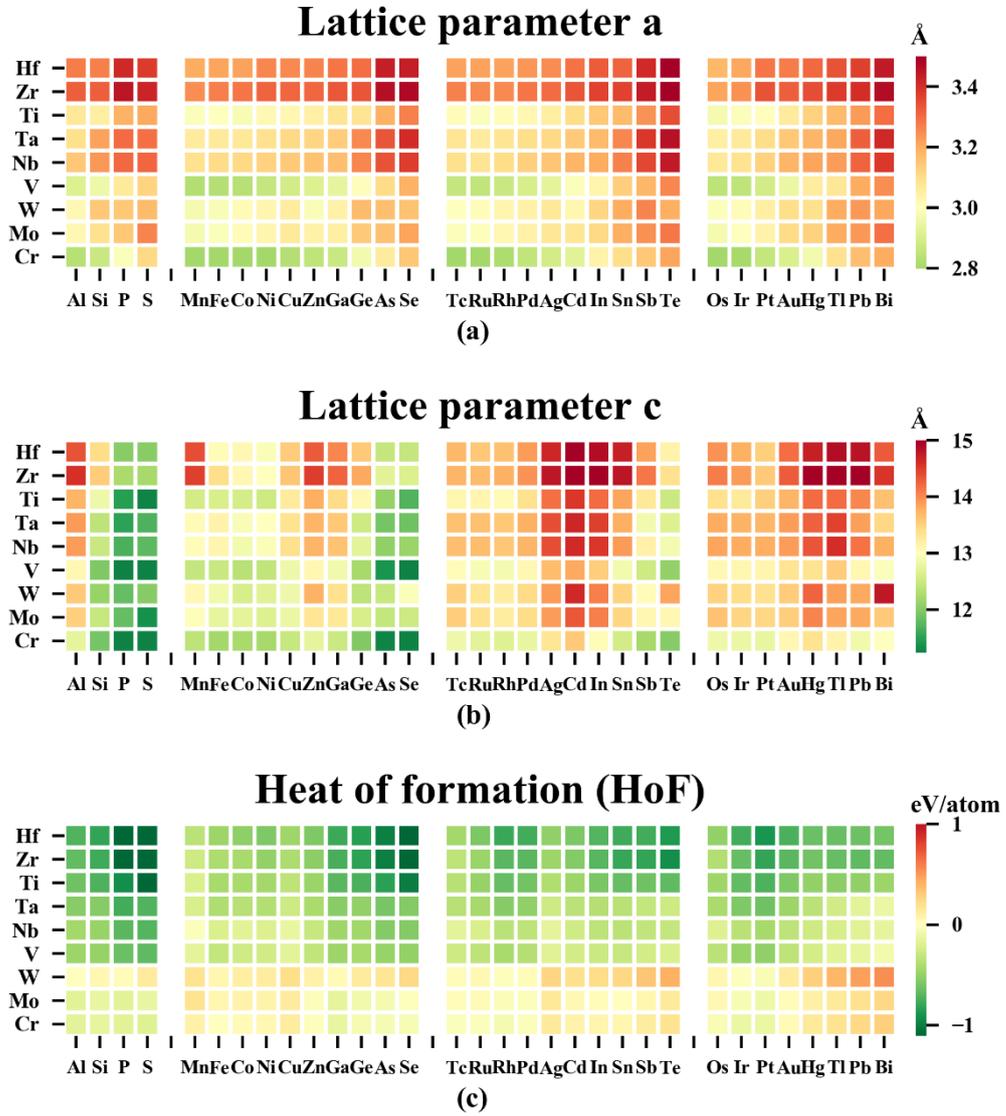

**Fig. 2.** (a) Heatmaps of lattice parameter in the basal plane $a$; (b) Heatmaps of lattice parameter along the $c$-axis $c$; (c) Heatmaps of heat of formation (HoF) for the MAX candidates, $E_{form}^{MAX}$.

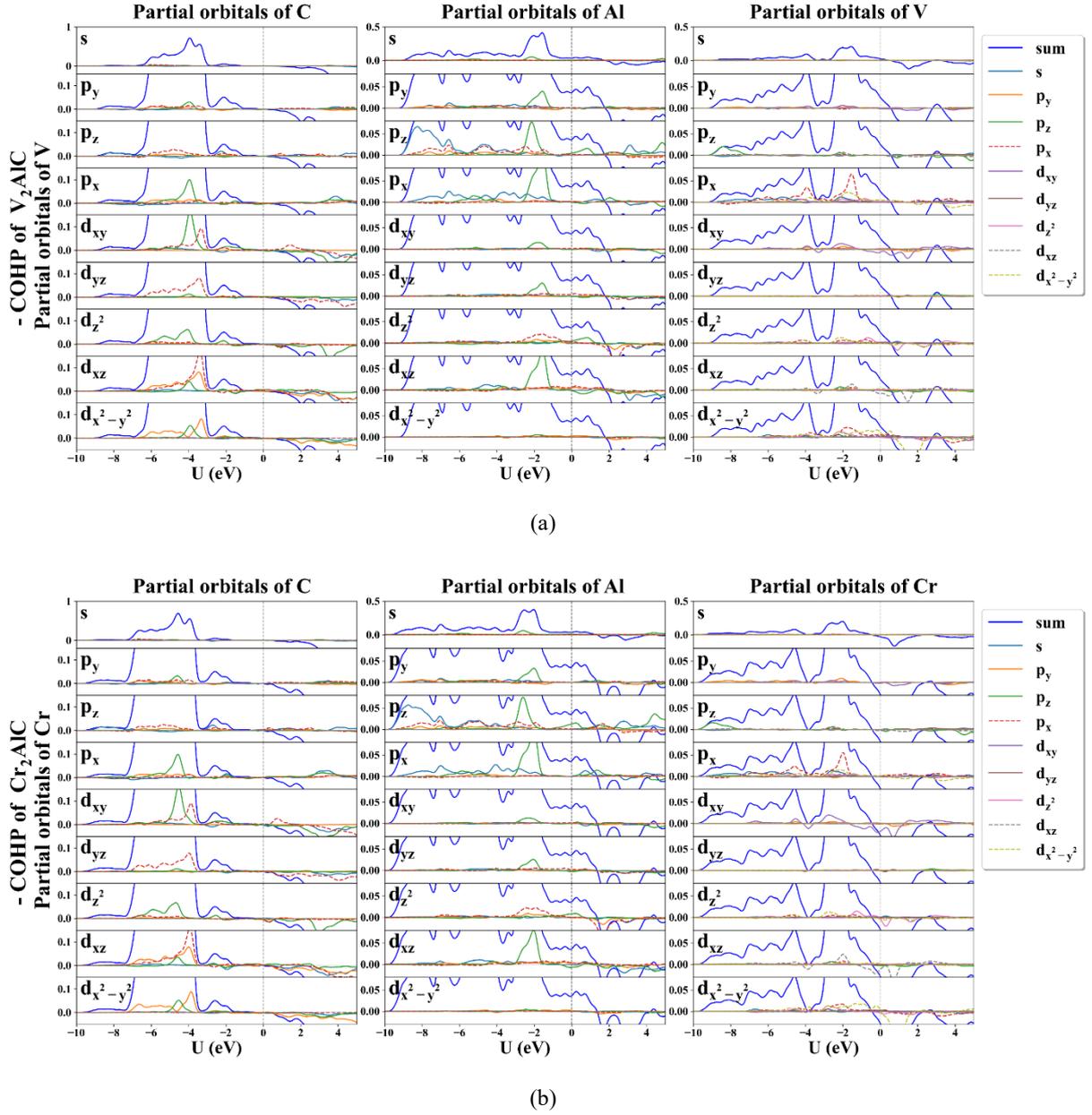

**Fig. 3.** The crystal overlap Hamilton population (COHP) analysis for (a) V$_2$AlC, (b) Cr$_2$AlC.

With this mechanism, it has been shown that the formability of MAX phases is attributed to the following factors: 1) VEC; 2) *geometrical* factor. For the VEC factor, its influence is mainly embodied in the ability to provide bonding electrons for materials, and the specific states of M, A hybrid bonding in different groups and periods. For the *geometrical* factor, it directly affects the solution energy required for A atoms to be embedded in the M-A-M twin layer. Under the premise of similar VEC, the difference of geometrical factors of different periodic elements is also one of the influencing factors of bonding or anti-bonding. To this end, we believe that describing the electronic structure and hybridization bonding laws of 288 M$_2$AX throughout the groups and periods is a critical step in unraveling the mystery.

For the law of elemental diversity of MAX Phases, we have explored clues by using the above consideration. Since the DOS diagram cannot provide complete bonding information, we use COHP to analyze the involved MAX phase in this work. For A-site elements of the same period, the increase of itinerant electrons causes the bonding state of M-A to move down to the Fermi level instead of M-X (Fig. S1); that is to say, we only need to study the bonding changes of M-A when M-site is determined. The bonding law of a series of MAX phases at the Fermi level has been discovered: 1) When the A-site elements are the groups of VIIB to IB, their *d* orbitals that lie between the *sp* hybrid state are involved in the bonding of M-A (Fig. S2 a-e); 2) When the A-site elements are the IIB group, the lagged *d* orbitals are only hybridized with *s* orbitals (Fig. S2f); 3) For A-site elements are in IIIA&IVA groups, the *d* orbitals fade out of the energy range of *sp* hybridization (Fig. S2 g-h); 4) As for A-site elements are in VA&VIA groups, the hybrid state of the *s* and *p* orbitals nearly disappears (Fig. S2 i-j). These provide apparent clues for understanding the nature of MAX phases.

Furthermore, when the A-site elements change from Group IIIA to Group VIA, M-A $d$-$p$ will reach saturation or supersaturation at Group VIA regardless of the VE of the M-site elements (Fig. S3). This implies that when VEC reaches 8.4, the anti-bond of M-M $d$-$d$ will still be intensified by adding of itinerant electrons of A-site elements, although their negative contribution is lower than the overload of valence electrons of M-site elements. When the transition metal A-site elements changes from VIIB to IIB, the bonding tendencies of MAX phases are dominated by the $VEC_M$. The increasing of itinerant electrons of A-site elements in the same period will increases the bonding electrons of M-A $d$-$p$ (Fig. S2I and S2II). Especially, the M-A (M=Cr, Mo, W) $d$-$p$ becomes saturated and the anti-bond of M-M $d$-$d$ becomes the main component of negative contribution (Fig. S2II a-f).

The effect of *geometrical* factor which due to the different atomic radii for different periods is discussed as follows. With the increase of the atomic size of MAX phases, especially in the a-axis direction (Fig. 2a), the bonding state of M-A $p\&d_{t2g}$-$p_z$ is dispersed; that is, the gradient of bonding state becomes smaller for the filling of electron energy level and the directionality of bonding is weakened (Fig. S4 I *vs* II). This leads to a lack of numerical gradients in the VEC characterization of the MAX (A= transition metal) phases at high periods.

3.2 The investigation of the variance of mechanical properties

In this chapter, the mechanical properties of MAX phases are studied, including: 1) bulk modulus (B), 2) shear modulus (G), 3) Young's modulus (E), 4) Poisson's ratio, 5) B/G and 6) representative independent modulus of elasticity $C_{11}$, $C_{33}$, $C_{44}$, as shown in Fig. 4 a-h. From the numerical point of view, the B/G (Poisson's ratio) of MAX (A= main groups) phases show the same rule as before: when VEC < 8.4 [45], the MAX phases show intrinsic brittleness; while greater than 8.4, MAX phases show the trend of low G which mainly due to the low $C_{44}$, showing potential as dry lubricant materials. In addition, based on Zhou *et al.* [11], we have observed the effects of $VEC_M$ on $C_{44}$ for different elemental period cases, wavelike dependence can be found for all $VEC_M$, although high $VEC_M$ leads to a generally decreasing trend with $VEC_A$ (Fig. S5a i-iii). For $C_{11}(C_{33})$, with the change of the number of itinerant electrons of A, there is a tendency to decrease (increase) first and then increase(decrease) in the case of low $VEC_M$ except for $C_{11}$ of M = Zr (Fig. S5 b,c); $C_{11}$ and $C_{33}$ are highly correlated, and these MAX phases tend to be isotropic and have a wider distribution as the $VEC_M$ increases (Fig. 5 a-c). As for the MAX (A = transition metal) phases, $C_{11}$ and $C_{33}$ show a monotonic decreasing trend with the itinerant electron of A-site and increasing trend with the valence electron of M-site. When M-site are in the 5$^{th}$ and 6$^{th}$ period, the values of $C_{11}$, $C_{33}$, B, E of MAX (A-site in VIIB and VIII groups) phases are larger than those with A=major groups on the whole. In addition, due to the partial absence of electrons in the bonding band of M-A $d$-$p$, the $C_{44}$ increases obviously (while M= Ti, V, Zr, Nb, Hf, Ta) with A, and the smaller G shows the ductility of M-A layer; when M = Cr, Mo, W, $C_{44}$ decrease with A. With the increase of M-site periods, it can be seen the gradient of increasing or decreasing gradually decreases in agreement with the effect of *geometrical* factors. The new trend of MAX (A = transition metal) phases is supposed to provide a new direction for the development of structural materials and dry lubricant materials.

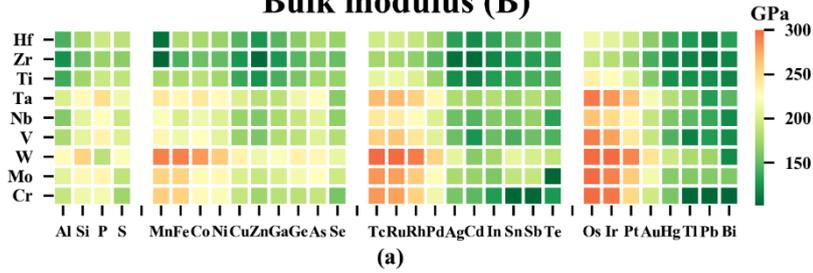
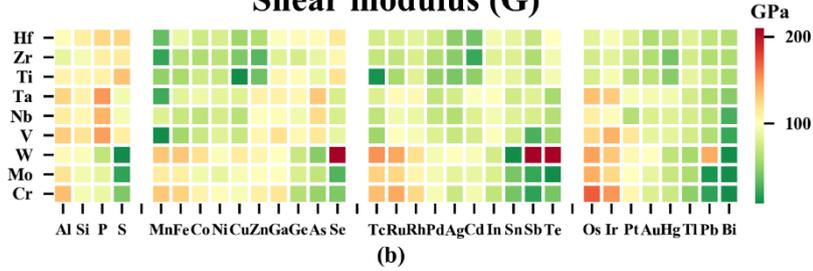
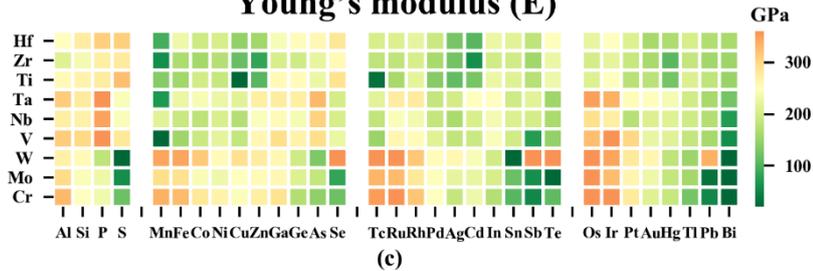
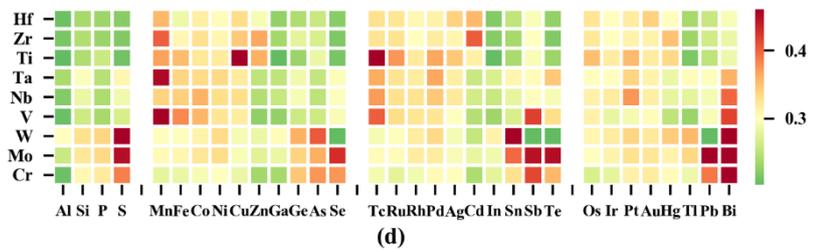
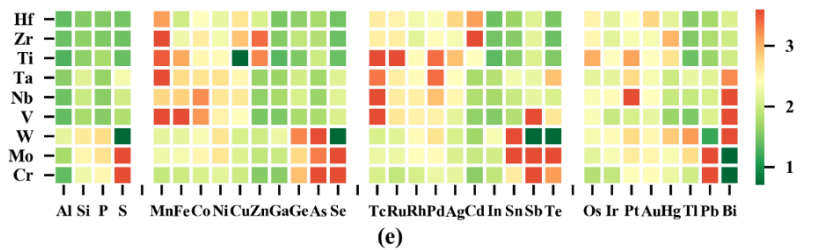

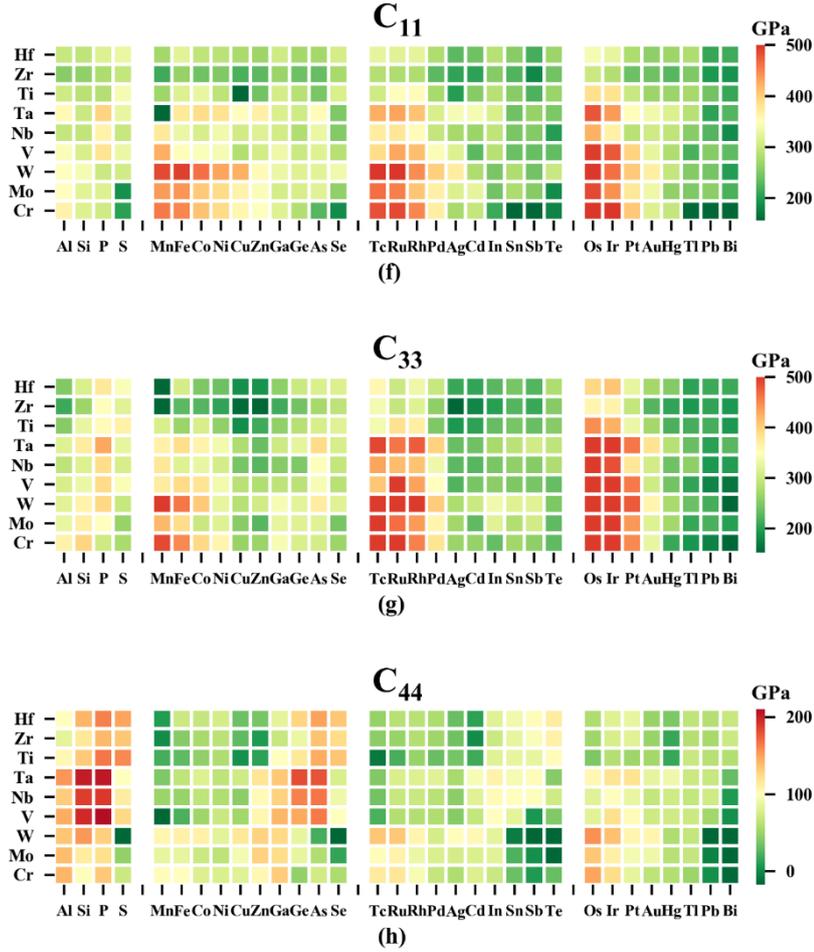

**Fig. 4.** (a) Heatmaps of bulk modulus, $B$; (b) Heatmaps of shear modulus, $G$; (c) Heatmaps of Young's modulus, $E$; (d) Heatmaps of Poisson's ratios, $v$; (e) Heatmaps of B/G values; (f) Heatmaps of elastic constant $C_{11}$ values; (g) Heatmaps of elastic constant $C_{33}$ values; and (h) Heatmaps of elastic constant $C_{44}$ values.

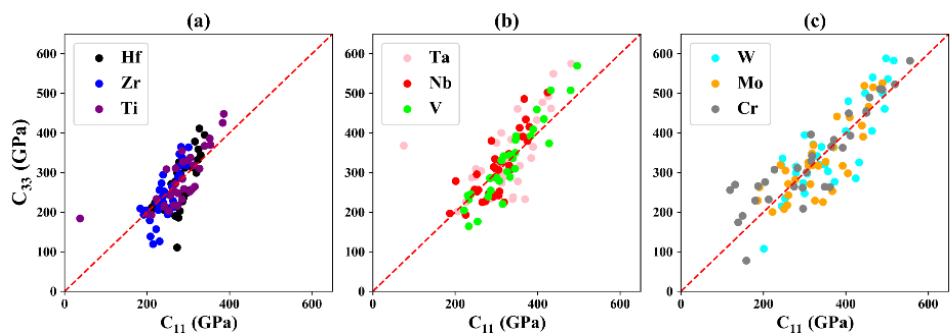

**Fig. 5.** The correlation between elastic constants $C_{11}$ and elastic constant $C_{33}$ in different $VE_M$ that (a) M = Hf, Zr, Ti, (b) M = Ta, Nb, V, (c) M = W, Mo, Cr.

As mentioned in the previous section, the mechanical properties of MAX phases as a transition metal carbide and nitride are significantly modulated by VEC. Meanwhile, the characteristics of electronic properties including TBOD, ICOHP, bond order and electronegativity have shown correlations with the force constants of MAX phases [24, 43, 45]. In this work, the COHP also played a key role in the analysis of mechanical performance trends.

For MAX (A-site in IIIA group) phases, as for many transition metal carbides [46], VEC = 8.4 $e$ is the dividing line between M-A $d$-$p$ bonding and M-M $d$-$d$ antibonding, showing a trend of increasing hardness first and then decreasing hardness [11, 47]. Due to the lack of flexible slip system [24, 47], the hardness of MAX phases is often related to the elastic constant $C_{44}$. However, while the element selectivity of M-site and A-site (main groups) becoming widespread, we have to consider the restriction of M-A $d$-$p$ antibonding. As mentioned in the previous section, when VEC > 8.4 for MAX (M = Cr, Mo, W; A = main groups) phases, the negative contribution of M-M $d$-$d$ anti-bond will be intensified leading to the decrease of $C_{44}$ in the traditional MAX phases even when the number of itinerant electrons of A-site increases (whatever the M-A $d$-$p$ bonding contribution is positive or negative). It means that the negative contribution of M-M $d$-$d$ anti-bond is dominant. Since the M-A (A= main groups) $d$-$p$ antibonding always occurs with A-site elements about in VA group, most $C_{44}$ of MAX phases will decrease in this case. For MAX (M = Ti, Zr, Hf; A = In, Sn, Sb, Te) phases, the decrease of $C_{44}$ may be not observed because M-site elements can provide enough states for electrons. From these COHP (Fig. S3 II - (a-b) instead of II - (c-f)), we observe that only on these periods, the M-A $d$-$p$ anti-bonds have not been filled, as an alternative perspective. This may be attributed to the larger size of the A-site atom, where the requirement for the electron concentration of fully bonded lightly increasing, resulting in the interesting phenomenon that the effect of the *geometrical* factor leads to a shift in the trend.

As for the MAX (M-site in fourth period; A= transition metal) phases, the $C_{44}$ still satisfies the rule of M-a $p\&d_{t2g}$-$p_z$, that is, M-A bonding increases with Ti and V, but saturates and antibonding increases with Cr. And in the case of M-site elements for the sixth period, $C_{44}$ does not show the same trend in elastic constants as they did in the fourth period. This may be explained by the dispersion behavior of M-A $p\&d_{t2g}$-$p_z$ bonding as the atomic mentioned above increasing.

Considered the weakening of the $C_{44}$ characteristic gradient, for the MAX (A= transition metal) phases with alloyed M-A-M layers, the hardness should be evaluated jointly with the role of $C_{11}$ and $C_{33}$. Finally, the ICOPH intensity of M-A $d$-$d$ is found to be significantly positively correlated with $C_{33}$ (Fig. 6). Therefore, the valence electron concentration, the *geometrical* factor and the binding strength of M-A $d$-$d$ need to be considered simultaneously in the design of the material.

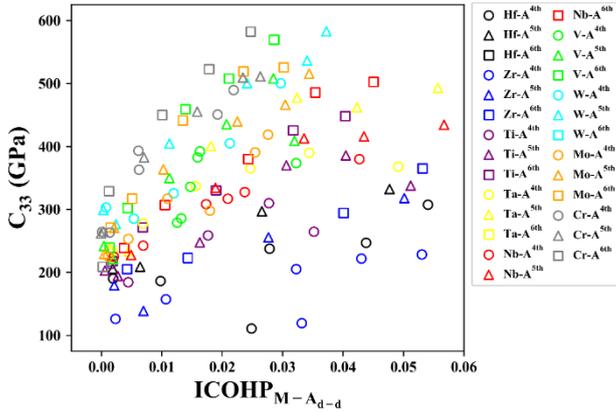

**Fig. 6.** The relationship between elastic constants $C_{33}$ and calculated ICOHP of M-A$_{d-d}$.

4. **Summary**

In summary, through a comprehensive theoretical study of the stability of 288 MAX phases, we have shown that the cohesion and structural stability could be characterized by the *electron concentration* (to evaluate the bonding energy) and factor (to evaluate the expansion energy). We have studied the electronic structure of the MAX phases as the A-site shift from VIIB to VIA groups and classified them into four main hybridization bonding states. Importantly, within the constraints of the rigid band model, the filling state of the bonding electrons of the M-site (A-site) elements more strongly (lightly) influence the formation of bonding or antibonding of M-A $d$-$p$ and M-M $d$-$d$, determining the effect of the continuous increase of valence electrons within the same period on the mechanical properties of MAX phases (especially on $C_{44}$ and G). This directly leads to the overall low shear modulus of MAX phases at low VEC (M in IVB group, A in VIIB and VIII groups) or hight VEC (M in VIB group, A in VA and VIA groups) and their potential as dry lubricant materials; while MAX (M in VIB group, A in VIIB and VIII groups) phases are predicted to have more appropriate B and G which may have both ceramic and metallic double-optimal properties due to the saturation of M-A $d$-$p$ bonding. Whereas, as the elemental period increases (the increase in atomic radius leads to a difference in the *geometrical* factor), the bonding effect and orientation of these bonds are weakened, leading to a decrease in the efficiency of the effect of bonding electron on the mechanical properties. For MAX phases where the A-site element is transition metal, the $C_{33}$ of MAX phases show a negative correlation with the arrangement of A-site element within the same period due to the addition of the A-site $d$-electron; in particular, values ($C_{33}$, B) of MAX (M=M in VIB group, A in VIIB and VIII groups) phases are much higher than those of other MAX phases within the same period due to the more significant M-A $d$-$d$ interactions. This study summarizes the overall trends and characteristics of the stability, electronic and mechanical properties of MAX phases, and reveals the mechanism how the electronic structure of MAX phases affects their mechanical properties from different perspectives.

**Acknowledgments**

The authors acknowledge the financial support of the National Key Research and Development Program of China (No. 2016YFB0700100, 2019YFB1901001), the Zhejiang Province Key Research and Development Program (No. 2019C01060), National Natural Science Foundation of China (Grants No. 21875271, 21707147, 11604346, 21671195, 51872302), K.C.Wong Education Foundation (rczx0800), the Foundation of State Key Laboratory of Coal Conversion (Grant J18-19-301), the project of the key technology for virtue reactors from NPIC, and the defense industrial technology development program JCKY 2017201C016.

**Supporting Information**

The Rendering from the Periodic System of the Elements on the stability, elastic, and electronic properties of $M_2AC$ phases

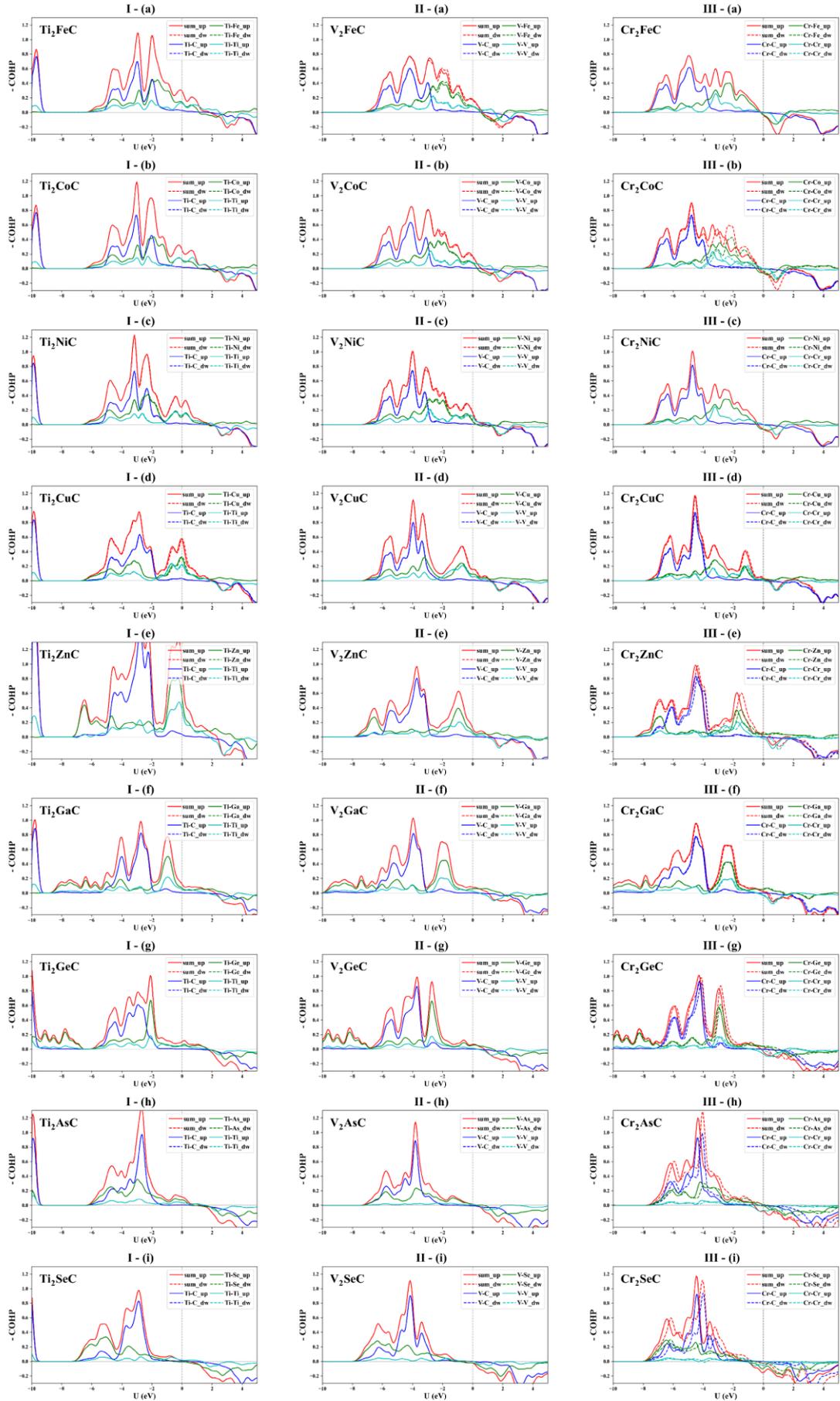

**Fig. S1.** The crystal orbital Hamilton population (COHP) of the main bonds (blue present M-X and green present M-A) in $M_2AC$ (M = (I) Ti, (II) V, (III) Cr, A = (a) Fe, (b) Co, (c) Ni, (d) Cu, (e) Zn, (f) Ga, (g) Ge, (h) As, (i) Se), where the solid and dashed lines respectively present the COHP for spin-up and spin-down orbitals.

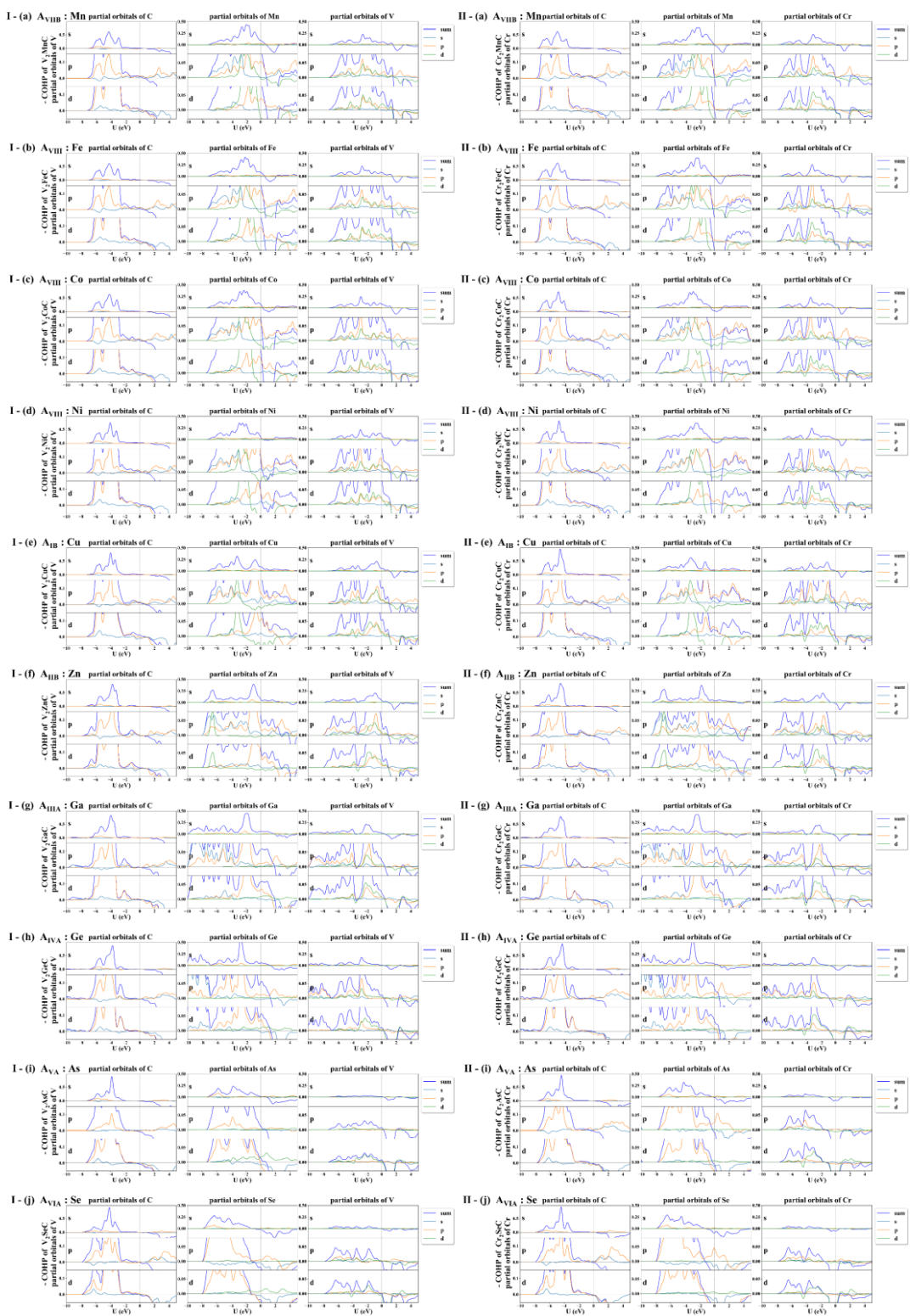

**Fig. S2** The crystal orbital Hamilton population (COHP) of the main orbital of bonds in M$_2$AC (M = (I) V, (II) Cr, A = (a) Mn, (b) Fe, (c) Co, (d) Ni, (e) Cu, (f) Zn, (g) Ga, (h) Ge, (i) As, (j) Se).

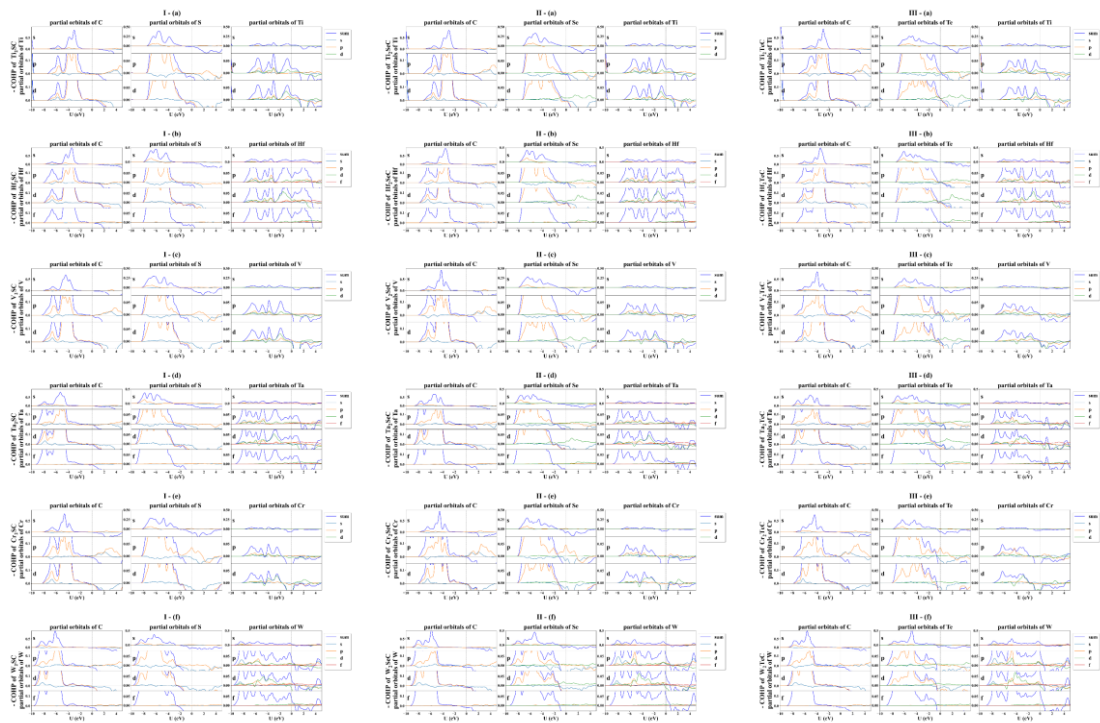

**Fig. S3.** The crystal orbital Hamilton population (COHP) of the main orbital of bonds in M$_2$AC (A = (I) S, (II) Se, (III) Te, M = (a) Ti, (b) Hf, (c) V, (d) Ta, (e) Cr, (f) W.

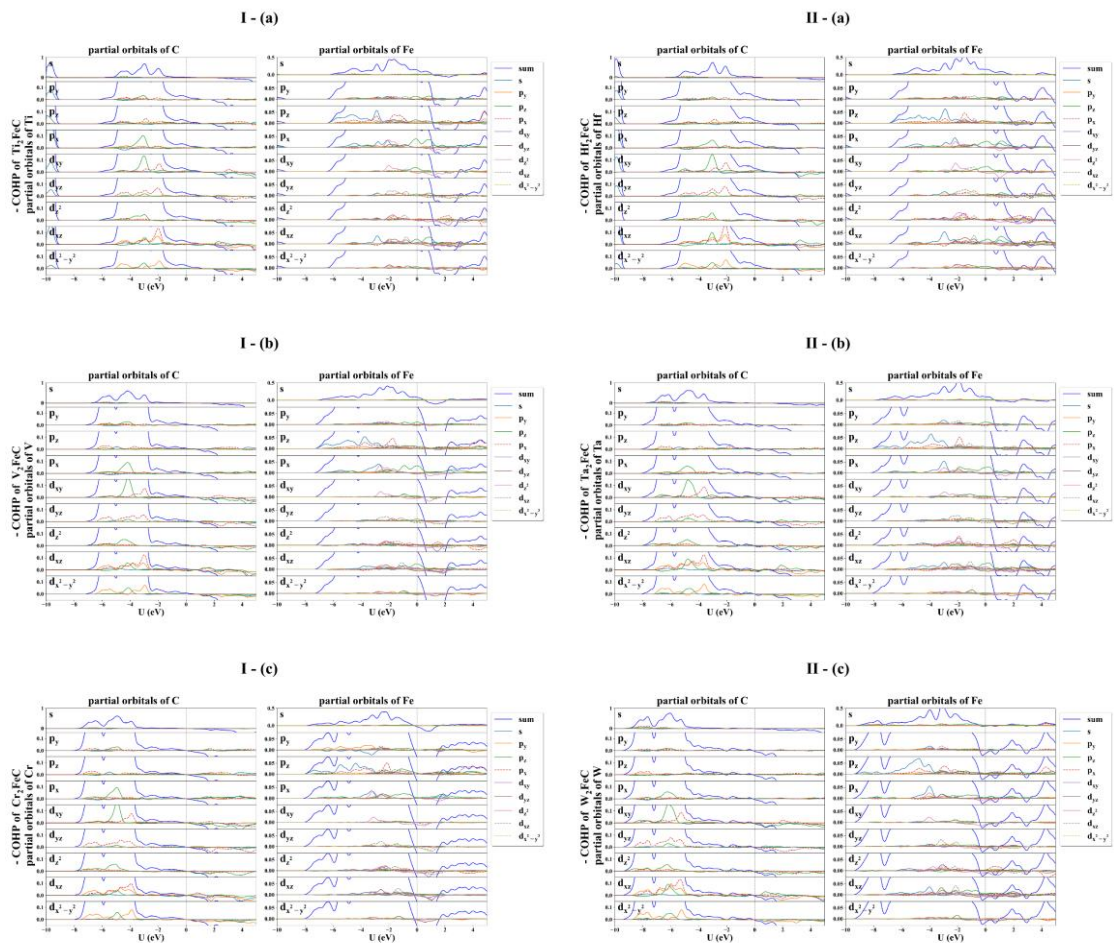

**Fig. S4.** The crystal orbital Hamilton population (COHP) of the partial orbital of bonds in $M_2FeC$ (M = I - (a) Ti, I- (b) V, I - (c) Cr, II - (d) Hf, II - (e) Ta, II - (f) W, where I (II) presents M-site elements in fourth (sixth) period.

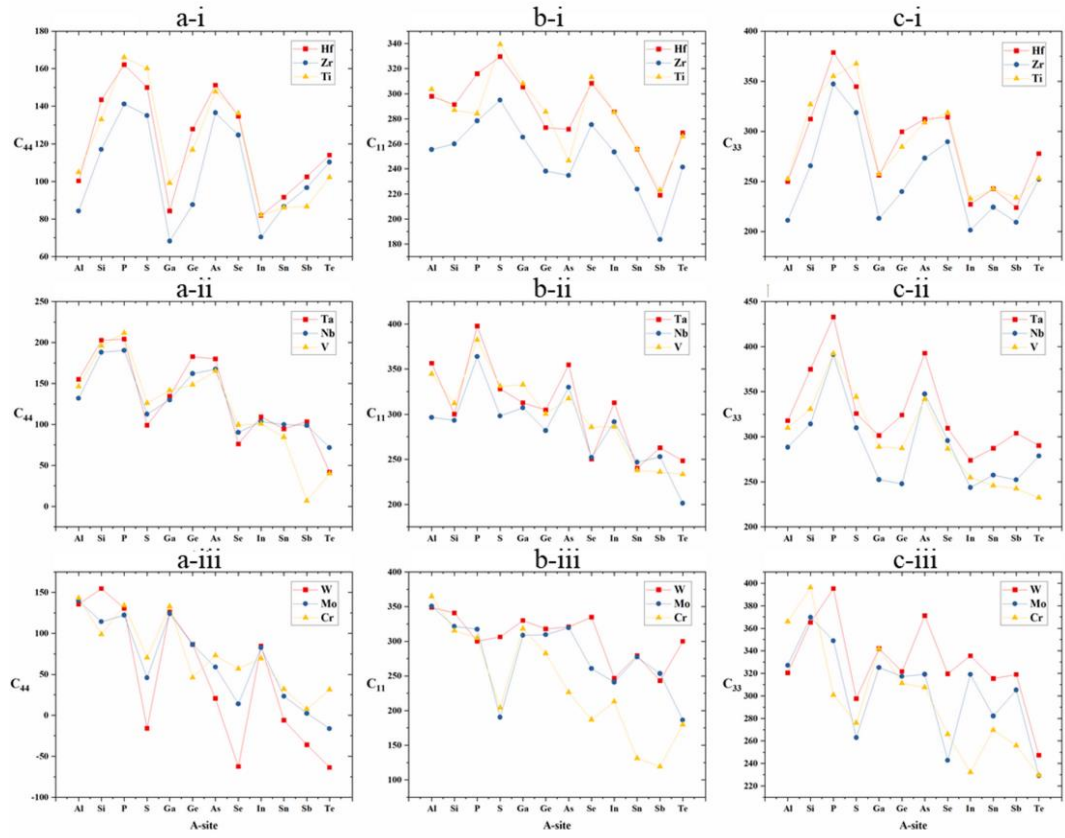

**Fig. S5.** Elastic constants $C_{44}$ (a), $C_{11}$ (b) and $C_{33}$ (c) of MAX (M= Ti, Zr, Hf (i), M= V, Nb, Ta (ii), M= Cr, Mo, W (iii)) phases.